\documentclass{aa}
\usepackage{graphics}
\usepackage{times}
\begin{document}

\thesaurus{01(11.11.1; 11.19.2; 11.19.6)}
\title{Hidden Bars and Boxy Bulges}
\author{Michael R. Merrifield\inst1 \and Konrad Kuijken\inst2
\thanks{Visiting Scientist, 
Dept Theoretical Physics, University of the Basque Country.}}
\institute{School of Physics and Astronomy, University of
Nottingham, NG7 2RD, England
\and
Kapteyn Institute, P.O. Box 800, 9700 AV Groningen, The Netherlands}
\offprints{Konrad Kuijken, {\small{kuijken@astro.rug.nl}}, Kapteyn
Institute, P.O. Box 800, 9700 AV Groningen, The Netherlands.}
\date{Received ??; accepted ??}
\maketitle

\begin{abstract}

It has been suggested that the boxy and peanut-sha\-ped bulges found in
some edge-on galaxies are galactic bars viewed from the side.  We
investigate this hypothesis by presenting emission-line spectra for a
sample of 10 edge-on galaxies that display a variety of bulge
morphologies.  To avoid potential biases in the classification of
this morphology, we use an objective measure of bulge
shape.  Generally, bulges classified as more boxy show the more
complicated kinematics characteristic of edge-on bars, confirming the
intimate relation between the two phenomena.

\end{abstract}

\section{Introduction}

For some considerable time, N-Body simulations have predic\-ted the
existence of a vertical instability in galactic bars (Com\-bes \&
Sanders \cite{cs81}, Combes et al.\ \cite{cdfp90}, Raha et al.\
\cite{rea91}). These simulations show that a bar forming in a flat
disk will not remain thin, but will quickly buckle and form a
thickened structure perpendicular to the plane of the disk.  Viewed
edge-on, such fat bars have a characteristic shape with very boxy
isopho\-tes.  In the most extreme cases, the structure appears
double-lobed, like a peanut in its shell.  These simulations are
intriguing because observations of real edge-on galaxies reveal that a
significant fraction have central bulges with boxy or peanut-shaped
isophotes (Jarvis \cite{j86}, Shaw \cite{s87}, de Souza \& dos Anjos
\cite{dd87}).  It is therefore tempting to associate such bulge
morphologies with edge-on bars, and even draw more general conclusions
regarding the formation of all bulges.

Unfortunately, establishing the link between bars and boxy bul\-ges
observationally has proved difficult.  Bars are recognizable only in
fairly face-on galaxies, while the vertical structure of a bulge can
only be deduced from an edge-on view.  Thus, it has proved impossible
to connect the two phenomena unequivocally using photometric data.

A few years ago, we suggested that bars could be detected in edge-on
galaxies from their kinematic signature (Kuijken \& Merrifield
\cite{km95}).  The orbits followed by material in a barred potential
will be non-circular, and the arrangement of the orbits changes
abruptly near resonances.  As a consequence of this complexity, the
observable line-of-sight velocities of material as a function of
position in an edge-on barred galaxy will also display complex
structure, with multiple components and gaps associated with the
resonances.  The existence of this complexity, which we originally
investigated using simple perturbation theory applied to closed
orbits, has subsequently been confirmed by full hydrodynamical gas
simulations (Athanassoula \& Bureau \cite{ba99}).

In a pilot spectral study, we found such kinematic structure in both
the stellar and gaseous components of two edge-on disk galaxies
(Kuijken \& Merrifield \cite{km95}).  Since these galaxies were
selected because they contained peanut-shaped bulges, this discovery
provided some evidence that bars and boxy bul\-ges are the same
phenomenon viewed from different directions.  However, the sample size
was very small, and lacked comparison data from galaxies without boxy
bulges.

We have therefore now carried out a larger spectral survey of edge-on
galaxies.  The sample was taken from the largest early-type disk
galaxies observable from the northern hemisphere in the RC3 catalog
(de Vaucouleurs et al.\ \cite{dvea91}), from which we selected a
subset of 10 galaxies designed to span a complete range in bulge
morphology, from elliptical to peanut-shaped.  Section~\ref{shapesec}
describes how the shapes of these bulges have been quantified, while
Sect.~\ref{specsec} presents the spectral data.  The results and their
implications are discussed in Sect.~\ref{concsec}.

\section{Bulge shapes}\label{shapesec}

Previous studies of bulge shapes (e.g.\ Jarvis \cite{j86}, Shaw
\cite{s87}, de Souza \& dos Anjos \cite{dd87}) have been based on the
visual impression of galaxies in sky survey plates.  Unfortunately,
this subjective approach cannot be used reliably when comparing the
bulge shapes to other properties of galaxies such as their
kinematics.  If, for example, we were to detect complex kinematics in
a galaxy, there is a significant risk that we would then reinforce out
initial prejudice by convincing ourselves that there were signs of
boxiness in the galaxy's isophotes.  What we require, therefore, is
some more objective approach.  

In the case of elliptical galaxies, shapes are relatively easy to
classify by measuring the minor departures of the isophotes from
ellipses (e.g.\ Bender, D\"obereiner \& M\"ollenhoff \cite{bdm88}).
However, the analysis of an edge-on disk galaxy is less
straightforward, as the contribution to the total light from the disk,
as well as dust absorption in the disk plane somewhat confuse the
bulge isophotal shapes. Nevertheless, after some experiments we have
found that it is possible to obtain robust measures of the bulge
isophotes using the techniques developed for elliptical galaxies.  

Images of most of our sample galaxies' bulges were obtained in the I
band at the William Herschel Telescope along with the spectral data
described below. Where this was not possible, we have searched the La
Palma archive, or failing this, used the Digitized Sky Survey (DSS).
We have measured the bulge isophotes with the ELLIPSE task in the
STSDAS analysis package of IRAF (Jedrjezewski \cite{j87}). We masked
out a wedge-shaped region of each image within 12 degrees of the disk
major axis to minimize the disk influence, and, in order to minimize
the effects of extinction by the disk, only fitted on the side of the
galaxy where the disk projects behind the bulge. The masking process
leaves less than half of the isophote available for fitting, but by
fixing the centroid and position angle of the isophotes to coincide
with those of the disk stable results can still be obtained.

\begin{figure*}
\resizebox{\hsize}{!}{\includegraphics{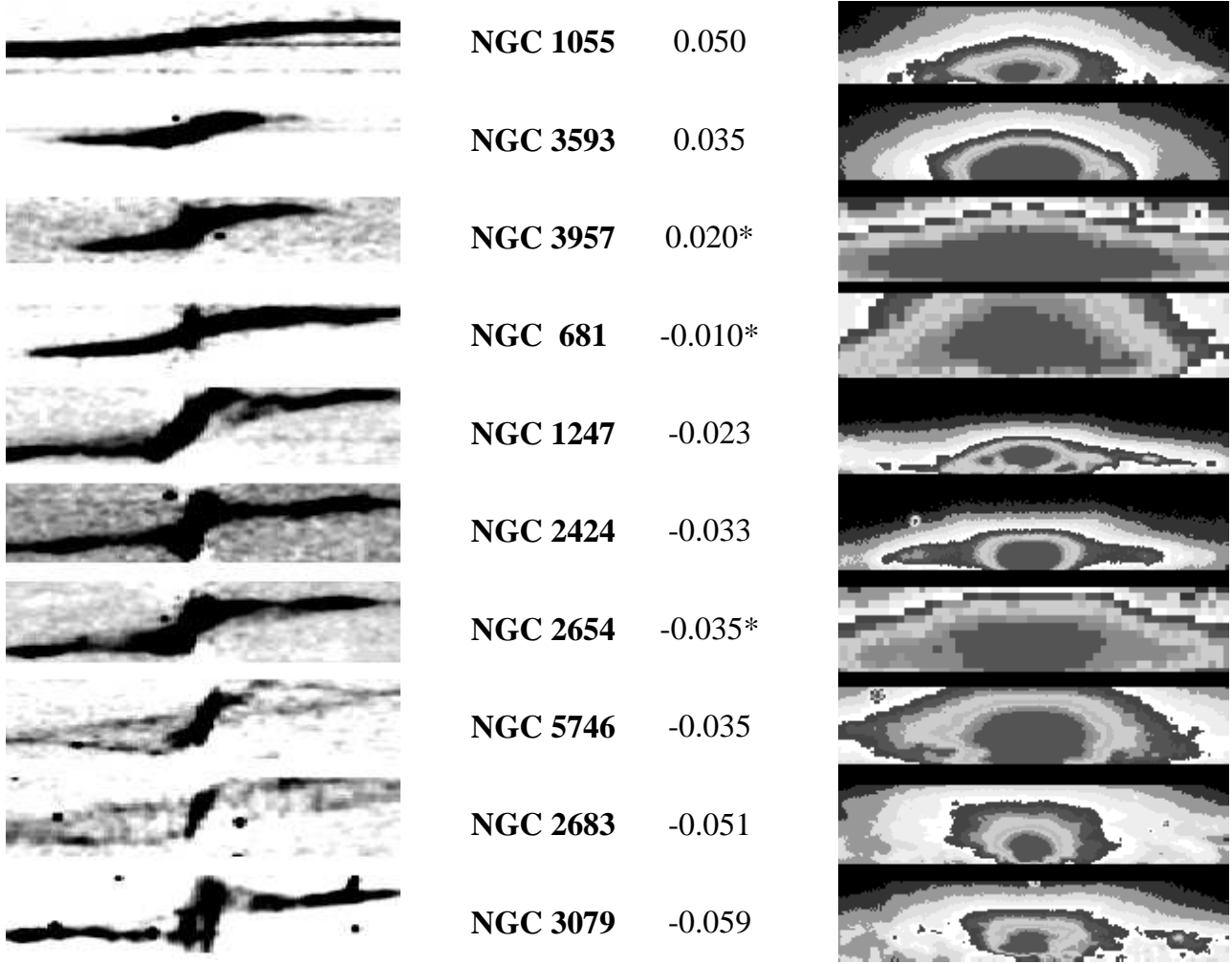}}
\caption{Montage of the sample of 10 galaxies sorted by 
deviation of their bulge isophotes from disky (top) to boxy (bottom).
The right panels
show I-band CCD images of the galaxies,
except for those marked with an asterisk which were taken from the Digitized
Sky Survey (DSS).  The left panels show the two-dimensional
[\ion{N}{ii}] emission line spectra, with position along major axis as
the $x$-axis and wavelength as the $y$-axis. All panels are 80 arcsecs
wide.
The $a_4$ isophote shape parameter for the least elliptical bulge
isophote is listed next to the galaxy name.}
\label{nutfig} 
\end{figure*}

Having measured the isophote shapes at a range of surface
brightnesses, we then classified a galaxy's boxiness on the basis of
the most extreme value (positive or negative) of the $a_4$ isophote
shape parameter.  Images of the half of the bulges to which the fit
was applied, ordered by the value of this parameter, are presented in
Fig.~\ref{nutfig}.  Reassuringly, this objective ordering process
arranges the galaxies in almost exactly the same sequence as one would
have done by eye, but without the dangers of {\it a posteriori} bias
that are inherent in any subjective classification.

\section{Spectral data}\label{specsec}

Using the ISIS spectrograph on the William Herschel $4.2\,{\rm m}$
telescope, we obtained long-slit spectra for all the galaxies in the
sample.  In each case, the spectrograph slit was aligned a\-long the
major axis of the galaxy (if necessary just avoiding the dust lane).
The spectra were obtained using a 1200-line grating, giving a
resolution of 1\AA\ (FWHM); the spectral range was centred on H$\alpha$
(6563\AA).  

We have previously found that the clearest signature of bar-induced
peculiar kinematics comes from the emission lines in the spectra
(Kuijken \& Merrifield \cite{km95}).  The strongest emission line in
this spectral region is H$\alpha$ itself, but it lies on top of the
H$\alpha$ absorption line associated with the stellar continuum, so
its signature is somewhat confused.  A clearer signal comes from the
neighbouring [\ion{N}{ii}] line at 6584\AA, which does not lie on top
of any significant absorption features.  We have therefore analyzed
the spectra in the vicinity of this line by first subtracting a
low-order fit to the stellar continuum, then subtracting sky spectra
from the ends of the slit so as to remove any night sky lines.  The
resulting two-dimensional spectra -- intensity of the [\ion{N}{ii}]
line as a function of position along the slit and wavelength -- are
shown in Fig.~\ref{nutfig}.  The wavelength scales in this figure have
been adjusted so as to present all the galaxies with similar
integrated line widths, thus rendering the kinematics from different
galaxies more directly comparable.

\section{Discussion}\label{concsec}

From inspection of Fig.~\ref{nutfig}, it is apparent that there is,
indeed, a link between bulge morphology and the complexity of the gas
kinematics -- generally speaking, galaxies with non-boxy bulges have a
simple kinematic structure, while boxier bulges seem to be associated
with complex multiple-component emission line kinematics.  Bureau and
Freeman (\cite{bf99}, \cite{b98}) reach a similar conclusion on the
basis of a sample of southern galaxies.  The structure apparent in the
more complex emission-line kinematics shown in Fig.~\ref{nutfig} --
X-shaped two-component systems, distorted central parallelogram
structures, and skewed figures-of-eight -- are exactly the classes of
feature that are generated by a non-axisymmetric barred potential when
viewed from a variety of angles (Merrifield \cite{m96}, Bureau \&
Athanassoula \cite{ba99}, Athanassoula \& Bureau \cite{ab99}).  It is
also notable that the radial scale over which the complex kinematics
occurs is comparable in extent to the bulges of the host systems,
again suggesting that the two phenomena are linked.

However, before we conclude from the association of boxy bulges with
complex kinematics that these systems are bars viewed edge-on, we
should consider other possible causes of complexity in the observed
kinematics.  One such possibility is that the gaps in the
emission-line profiles result from differential extinction by dust
down the line of sight.  However, in axisymmetric galaxies, the
line-of-sight velocity of the gas must be a smooth function of
distance down the line of sight, so it is difficult to see how partial
obscuration in an unbarred galaxy could result in multiple-components
in the kinematics.  

A second possibility is that the structure in the
kinematics arises from rings of gas in the galaxy, as are seen in a
number of more face-on disk systems (Buta \& Combes \cite{bc96}).  In
a two-dimensional spectrum of an edge-on galaxy, an axisymmetric ring
of emission will appear as an inclined straight line.  Superficially,
some of the structures in the spectra in Fig.~\ref{nutfig} conform to
this pattern.  However, there are several crucial differences.  First,
rings are edge-brightened when seen in projection, yet most of the
linear features in Fig.~\ref{nutfig} do not get brighter towards their
ends.  Second, axisymmetric rings project to straight lines that pass
through the systemic velocity of the galaxy at its centre, whereas
many of the linear features in Fig.~\ref{nutfig} are not quite
straight, and have non-zero velocities at the centres of their
galaxies.  Such features cannot occur in an axisymmetric potential, so
we are once again forced to conclude that these galaxies are barred.

A third possibility, finally, is that we are seeing variations in
ionization structure of the gas, and not inhomogeneities in the
distribution of the gas itself. In fact, in barred galaxies we also
expect such variations, induced by the shocks; and they are indeed
observed as systematic variations in N[II]/H$\alpha$ ratio over the
$(R,v)$ diagrams (see also Bureau \cite{b98}). A detailed analysis of
the line ratio lies beyond the scope of this paper.  However, as far
as can be ascertained, the H$\alpha$ emission line shows identical
structure to the [NII] line, providing further evidence that the
structure cannot be attributed to the details of the gas' ionization
state.

In summary, this spectral study of edge-on galaxies quite firmly
establishes the link between boxy bulges and galactic bars.  However,
we have only just begun to tap into the wealth of information that the
spectra provide.  Modelling the full complexity of spectral data such
as those shown in Fig.~\ref{nutfig} should yield a wealth of
information about barred galaxies, allowing us to map out their
complete three-dimensional structure for the first time.

\begin{acknowledgements}
The Digitized Sky Survey was produced by STScI under U.S. Government
grant NAG W-2166, and is subject to a variety of copyrights ({\tt
http://{\rm \-}stdatu.stsci.edu/{\rm \-}dss/dss\_copy{\rm \-}right.html}).
The William
Herschel Telescope is operated on the island of La Palma by the Isaac
Newton Group in the Spanish Observatorio del Roque de los Muchachos of
the Instituto de Astrofisica de Canarias. Much of the analysis in this
paper was performed using {\sc iraf}, which is distributed by NOAO,
operated by AURA under cooperative agreement with the NSF{}.  MRM has
been supported by a PPARC Advanced Fellowship (B/94/AF/1840)
\end{acknowledgements}

\end{document}